# Robustness of Majorana zero-energy state

Zheng-Chuan Wang

Department of Physics & CAS Center for Excellence in Topological Quantum Computation, The University of Chinese Academy of Sciences, P. O. Box 4588, Beijing 100049, China.

## Abstract

Based on the principle of linearized stability proposed by Lyapounov, we investigate the robustness of Majorana zero energy state (MZES), which plays an important role in topological quantum computation. We show that the MZES is not enough robust against the external perturbations, because mathematically it is critical stable instead of the asymptotic stable, only the states with asymptotic stability can be regarded as robustness, so the MZES can not be used to carry quantum information in topological quantum computation. Our study is different from previous works that usually make the numerical test by some special perturbations, our analytical derivation is suitable for arbitrary perturbations. As an example, we demonstrate it by the stability analysis of MZES in the spin-orbit coupled semiconductor/ superconductor junction.

## Introduction

In 1996, Shor presented a Fault-tolerant quantum computation[1] to overcome the difficulty caused by the decoherence of quantum bits. However, it is difficult to achieve due to its small threshold value for those errors in quantum logical gates, so some treatments of topological quantum computation[2,3] were proposed to avoid the problem of decoherence, in which a decoherence free subspace is exploited to process the quantum information. The decoherence free subspace of degenerate states can exist in some two-dimensional system, such as chiral p-wave superconductor[4], the $\nu = \frac{5}{2}$ fractional quantum Hall system[5], and some heterostructures, i.e., s-wave superconductor /topological insulator[6], spin-orbit coupled semiconductor/ superconductor system[7], where the Majorana bound states-- the elementary excitations obeying the non-Abelian statistics can be used to carry quantum information, it is robust against the noise.

Topological stability of MZES had been investigated extensively[8], especially in superconductor/topological system[9], s-wave superconductivity on the surface of a spin-orbit coupled semiconductor [10,11], and superconductor/ferromagnet junction on a strong topological insulator surface[12]. However, S. Das Sarma et al. pointed out that the stability of MZES can be destroyed by the quantum tunneling process of Majorana Fermion between vortex[13-15], in which the ground state degeneracy is split. Cheng et al. calculated the splitting of the zero-energy ground state in a two-dimensional p-wave superconductor topological insulator/superconductor structure and superconductor/semiconductor hybrid topological system[13-14]. An experiment in InAs nanowire segment with epitaxial aluminum had observed this splitting[16]. These works tell us that the robustness of MZES is worthwhile to be further explored.

Since the bulk gap can protect the ground state degeneracy from perturbation and keep it's stability, people usually concentrate on the calculation of gap of quasiparticle excitation in the vortex[10,11]. There exist non-topological Caroli-de Gennes-Matricon (CdGM) states localized inside the core of vortex on the MZES, the minigap between the Majorana modes and the band of single particle non-topological excitation is very small, i.e., in the spin-orbit coupled semiconductor/ superconductor junction, the MZES may be easily disturbed by the thermal fluctuations. Despite of the argument that the MZES is still robust by some authors, it is worthwhile to reexamine the stability of MZES by other ways.

As we know, people often discuss the topological stability of the system by investigating the topological characters of energy bands or the phase transition, they seldom directly discuss the robustness of wavefunction. Since the quantum information is carried by the wavefunction, i.e., the MZES in the system, and the wavefunction is easily influenced by the decoherence, so we had better explore the stability by the wavefunction itself instead of the minigap. It should be noted that the MZES is a solution of Bogoliubov-de Gennes equation which is a differential equation. Mathematically, it is concerning with the stability analysis on the solution of a differential equation, whose theory had been well established by Lyapounov[17,18]. In this manuscript, we will investigate the stability of MZES by the method of linear stability analysis developed by Lyapounov[17,18]. Under this stability analysis, we find that the MZES is not enough robust because it is critical stable instead of asymptotic stable, it can't be exploited to carry the quantum information. We will show this by an example of spin-orbit coupled semiconductor/superconductor junction.

**Theoretical formalism**

Consider a system with Majorana bound states, such as the s-wave superconductor/topological insulator or the semiconductor with strong spin orbital coupling in proximity to superconductors, the MZES can occur on the interfaces of the above systems, which can be described by the following Bogoliubov de-Gennes Hamiltonian

$$\widehat{H} = \begin{pmatrix} \hat{h} & \Delta_{SC} \\ \Delta_{SC}^+ & -\hat{h} \end{pmatrix}, \qquad (1)$$

where the kinetic energy, potential, spin orbital coupling as well as Zeeman term are all included in the single particle Hamiltonian $\hat{h}$, and

$\Delta_{SC}$ represents the superconductor pair potential. The requirement of particle-hole symmetry for the Majorana bound states demands that the wavefunction must adopt the Nambu notation $\psi = ((\psi_\uparrow, \psi_\downarrow), (\psi_\downarrow^\dagger, -\psi_\uparrow^\dagger))$. According to the Bogoliubov de-Gennes eigen-equation $\hat{H}|\psi> = E|\psi>$, the MZES satisfies

$$\hat{H}|\psi>_{MZES} = 0, \qquad (2)$$

which corresponds to the zero energy $E = 0$ and $i\hbar \frac{\partial |\psi>_{MZES}}{\partial t} = 0$. Mathematically, $\psi_{MZES}$ is just an equilibrium solution for the time-dependent Bogoliubov de-Gennes equation, we can study its robustness by the well established theory of stability given by Lyapounov[17,18].

Suppose the solution of time dependent Bogoliubov de-Gennes equation can be written as the following form

$$|\psi> = |\psi>_{MZES} + |\delta\psi>, \qquad (3)$$

where $|\delta\psi>$ is the external perturbation away from the equilibrium solution. Substituting it into the Bogoliubov de-Gennes equation $i\hbar \frac{\partial |\psi>}{\partial t} = \hat{H}|\psi>$, we obtain the time dependent equation for the perturbation

$$i\hbar \frac{\partial |\delta\psi>}{\partial t} = \hat{H}|\delta\psi>, \qquad (4)$$

which is a equation about $|\delta\psi>$. If we adopt the perturbation as

$$|\delta\psi(\vec{r}, t)> = |\delta\psi_0(\vec{r})> e^{\lambda t}, \qquad (5)$$

where $|\delta\psi_0(\vec{r})> = ((\delta\psi_{\uparrow 0}(\vec{r}), \delta\psi_{\downarrow 0}(\vec{r})), (\delta\psi_{\downarrow 0}^\dagger(\vec{r}), -\delta\psi_{\uparrow 0}^\dagger(\vec{r})))^T$ describes the position dependence of perturbation. Inserting Eq.(5) into Eq.(4), we have

$$i\hbar\lambda|\delta\psi_0(\vec{r})> = \hat{H}|\delta\psi_0(\vec{r})>, \tag{6}$$

then the eigen-value $\lambda$ can be expressed as:

$$\lambda = \frac{<\delta\psi_0(\vec{r})|\hat{H}|\delta\psi_0(\vec{r})>}{i\hbar<\delta\psi_0(\vec{r})|\delta\psi_0(\vec{r})>}, \tag{7}$$

which is a pure imaginary number. To see this clearly, we can further expand $|\delta\psi_0(\vec{r})>$ by the eigen functions of Hamiltonian as: $|\delta\psi_0(\vec{r})> = \sum_n c_n|\psi_n(\vec{r})>$, where $\hat{H}|\psi_n(\vec{r})> = E_n|\psi_n(\vec{r})>$, and $E_n$ is the eigen energy, then $\lambda$ may be finally expressed as $\lambda = \frac{\sum_n |c_n|^2 E_n}{i\hbar \sum_n |c_n|^2}$, obviously, it is pure imaginary.

According to the principle of linearized stability proposed by Lyapounov[17,18], if all the eigen-values $\lambda$ have negative real parts, the $\psi_{MZES}$ is a asymptotic stable equilibrium solution; if some eigen-values $\lambda$ have positive real parts, then $\psi_{MZES}$ is unstable; if the real part of eigen value is zero, which means it is pure imaginary like Eq.(7), $\psi_{MZES}$ is called critical stable. Although both critical stability and asymptotic stability can be attributed to the Lyapounov stability, the critical stability is not robust as asymptotic stability. Only the asymptotic stable equilibrium solution is robust against any perturbations, while the critical stable solution is not enough robust[17,18], because it is inevitably determined by the initial conditions which are easily disturbed by the noise. People usually look for the physical states with asymptotic stability for realistic application, instead of those states with critical stability[17,18]. So the MZES $\psi_{MZES}$ with critical stability is not suitable to carry the quantum information, which is not enough robust against the perturbations. The MZES can't be regarded as a good candidate for topological quantum computation, we should search for other candidates with asymptotic stability. All in all we can judge the robustness of MZES $\psi_{MZES}$ by the principle of linearized stability instead of the minigap calculated by prior works. Since our method follows the stability theory of differential equation, the results we obtained are reliable, which is based on strict mathematical derivations. We will demonstrate this by an example of spin-orbit coupled semiconductor/superconductor junction in the next section.

**Example**

It is shown that the interaction between s-wave superconductor and spin-orbit coupled semiconductor by the proximity effect can induce the superconductivity in the topological surface states, so we consider a s-wave superconductor/spin-orbit coupled semiconductor/magnetic insulator junction as shown in Fig.1, it can be described by the following Hamiltonian

$$\widehat{H} = \begin{pmatrix} H_0 & \Delta_{SC}(r) \\ \Delta_{SC}^+(r) & -\sigma_y H_0^\dagger \sigma_y \end{pmatrix}, \tag{8}$$

where

$$H_0 = \begin{pmatrix} \frac{p^2}{2m^*} + V_z - \mu & i\alpha p_x + \alpha p_y \\ -i\alpha p_x + \alpha p_y & \frac{p^2}{2m^*} - V_z - \mu \end{pmatrix}, \tag{9}$$

in which $V_z$ denotes the perpendicular Zeeman field induced by the proximity contact with the magnetic insulator, $\mu$ is the chemical potential, and $\alpha$ is the strength of Rashba spin orbital coupling in the semiconductor. As pointed out by Mao et al.[11], the MZES can exist at the interface of superconductor and spin-orbit coupled semiconductor.

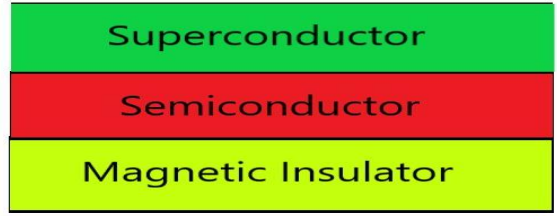

Fig.1 Schematic structure of a Superconductor/spin-orbit coupled Semiconductor/ Magnetic Insulator junction.

To demonstrate the stability of MZES in the s-wave superconductor/ spin-orbit coupled semiconductor/magnetic insulator junction, we should follow the same procedure as described in the above section. According to Eq.(6), the position part of perturbation $|\delta\psi_0(\vec{r})>= ((\delta\psi_{\uparrow 0}(\vec{r}), \delta\psi_{\downarrow 0}(\vec{r})), (\delta\psi_{\downarrow 0}^\dagger(\vec{r}), -\delta\psi_{\uparrow 0}^\dagger(\vec{r})))^T$ satisfies

$$i\hbar\lambda \begin{pmatrix} \begin{pmatrix} \delta\psi_{\uparrow 0}(\vec{r}) \\ \delta\psi_{\downarrow 0}(\vec{r}) \end{pmatrix} \\ \begin{pmatrix} \delta\psi_{\downarrow 0}^\dagger(\vec{r}) \\ -\delta\psi_{\uparrow 0}^\dagger(\vec{r}) \end{pmatrix} \end{pmatrix} = \begin{pmatrix} \frac{p^2}{2m^*}+V_z-\mu & i\alpha p_x+\alpha p_y & \Delta_{SC}(r) & 0 \\ -i\alpha p_x+\alpha p_y & \frac{p^2}{2m^*}-V_z-\mu & 0 & \Delta_{SC}(r) \\ \Delta_{SC}^*(r) & 0 & -\frac{p^2}{2m^*}+V_z+\mu & i\alpha p_x+\alpha p_y \\ 0 & \Delta_{SC}^*(r) & -i\alpha p_x+\alpha p_y & -\frac{p^2}{2m^*}-V_z+\mu \end{pmatrix} \begin{pmatrix} \begin{pmatrix} \delta\psi_{\uparrow 0}(\vec{r}) \\ \delta\psi_{\downarrow 0}(\vec{r}) \end{pmatrix} \\ \begin{pmatrix} \delta\psi_{\downarrow 0}^\dagger(\vec{r}) \\ -\delta\psi_{\uparrow 0}^\dagger(\vec{r}) \end{pmatrix} \end{pmatrix},$$

(10)

Pre-Multiplying $((\delta\psi_{\uparrow 0}^\dagger(\vec{r}), \delta\psi_{\downarrow 0}^\dagger(\vec{r})), (\delta\psi_{\downarrow 0}(\vec{r}), -\delta\psi_{\uparrow 0}(\vec{r})))$ on both sides of Eq.(10), we can obtain the eigen value $\lambda$ as

$$\lambda = \frac{A}{2i\hbar(|\delta\psi_{\uparrow 0}(\vec{r})|^2+|\delta\psi_{\downarrow 0}(\vec{r})|^2)} \tag{11}$$

where $A = 2Re[\delta\psi_{\uparrow 0}^\dagger(\vec{r})(i\alpha p_x + \alpha p_y)\delta\psi_{\downarrow 0}(\vec{r})] + 2Re\left(\Delta_{SC}(r)\delta\psi_{\uparrow 0}^\dagger(\vec{r})\delta\psi_{\downarrow 0}^\dagger(\vec{r})\right) - 2Re\left(\Delta_{SC}(r)\delta\psi_{\downarrow 0}(\vec{r})\delta\psi_{\uparrow 0}(\vec{r})\right) - 2Re[\delta\psi_{\downarrow 0}(\vec{r})(i\alpha p_x + \alpha p_y)\delta\psi_{\uparrow 0}^\dagger(\vec{r})] + \delta\psi_{\uparrow 0}^\dagger(\vec{r})\left(\frac{p^2}{2m^*} + V_z - \mu\right)\delta\psi_{\uparrow 0}(\vec{r}) - \delta\psi_{\uparrow 0}(\vec{r})\left(\frac{p^2}{2m^*} + V_z - \mu\right)\delta\psi_{\uparrow 0}^\dagger(\vec{r}) + \delta\psi_{\downarrow 0}^\dagger(\vec{r})(\frac{p^2}{2m^*} - V_z - \mu)\delta\psi_{\downarrow 0}(\vec{r}) - \delta\psi_{\downarrow 0}(\vec{r})(\frac{p^2}{2m^*} - V_z - \mu)\delta\psi_{\downarrow 0}^\dagger(\vec{r})]$, and $Re[...]$ stands for the real part of a term. The last four terms in the expression of $A$ are all real, because each of them is conjugate to itself, then $A$ is a real number and the eigen value $\lambda$ is pure imaginary.

So we conclude that the MZES in the s-wave superconductor/ spin-orbit coupled semiconductor/magnetic insulator junction is critical stable, it can't be used to carry the quantum information in topological quantum computation, because it is not enough robust against the perturbations according to the Lyapounov's principle of linearized stability[20,21]. We should look for the states with asymptotic stability in realistic systems to carry quantum bits. It should be pointed out that our results are strict, we haven't adopted any approximation in our derivations.

**Summary and discussion**

Although our conclusion in the above is different from the result presented by Mao et al.[11], it is conceded with the points given by Cheng et al.[12-14]. Mao's work only investigated several special perturbations during their numerical test, such as the spin-independent Gaussian impurity, a magnetic impurity potential $U(r)\sigma_z$, and a random impurity potentials, the verification of robustness for MZES in their work inevitably depends on these special perturbations. We can't discuss the stability of MZES only by this kind of numerical test with special perturbations. It is necessary to study the stability of MZES by other methods, such as the stability theory of differential equation developed by Lyapounov[17,18], it is a more general way for arbitrary perturbations.

We present a strict derivation to investigate the stability of MZES. It is shown that the MZES is critical stable, it is not enough robust against the external arbitrary perturbations, which means it can't be used to carry quantum information in topological computation. We should search for other physical states with asymptotic stability for quantum computation, which are robust in reality.

## Acknowledgments

This study is supported by the National Key R&D Program of China (Grant No. 2018FYA0305804), the Key Research Program of the Chinese Academy of Sciences (Grant No. XDPB08-3), and the Fundamental Research Funds for the Central Universities.